\documentclass[prb,aps,twocolumn,showpacs]{revtex4-1}

\usepackage{graphicx}

\usepackage{amsmath}
\usepackage[dvips]{color}
\usepackage{amssymb}
\usepackage{color}

\usepackage{bm}

\def\Tr{\mbox{Tr}\,}
\def\Im{\mbox{Im}}

\newcommand{\half}{\mbox{\small $\frac{1}{2}$}}
\newcommand{\eexp}{\mbox{e}^}

\begin{document}
\title{Nanoscopic interferometer model for spin resonance in current noise }
\author{Anatoly Golub and Baruch Horovitz }
\affiliation{Department of Physics, Ben-Gurion University of the Negev Beer-Sheva, Israel}
\begin{abstract}
 We study a model for the observed phenomenon of  electron spin resonance (ESR) at the Zeeman frequency as seen by a scanning tunneling microscope (STM) via its current noise.
The model for this ESR-STM phenomenon allows the STM current to flow in two arms of a nanoscopic interferometer, one arm has direct tunneling from the tip to the substrate while the second arm has tunneling through two spin states. We evaluate analytically the noise spectrum for non-polarized leads, as relevant to the experimental setup. We
show that spin-orbit interactions allow for an interference of two tunneling paths resulting in a resonance effect.
\end{abstract}
\maketitle
\section{Introduction}

 The control and detection of single spins is of considerable recent interest. A particularly interesting method of detecting a single spin on a surface is possible by a Scanning Tunneling Microscope (STM) \cite{review}. The technique has been initiated and developed by Y. Manassen and various collaborators \cite{review,manassen1,manassen2,manassen3}. It is based on monitoring the noise, i.e. the STM current-current correlations, and observing a signal at the expected Larmor frequency, a signal that is sharp even at room temperature. The Larmor frequency is also seen in an Electron Spin Resonance (ESR) experiment with many spins, in contrast, the ESR-STM method observes a single spin and furthermore, the system is static, no oscillating field is applied as in ESR. The observed frequency is found to vary linearly with the applied magnetic field, confirming that the STM has detected an isolated spin on the surface. This phenomenon was first demonstrated on oxidized Si(111) surface \cite{manassen1,manassen2} and then on Fe atoms \cite{manassen3} on Si(111) as well as on a variety of organic molecules on a graphite surface \cite{durkan} and on Au(111) surfaces \cite{messina,mannini,mugnaini}. Recent extensions have resolved two resonance peaks on oxidized Si(111) $7\times 7$ surface corresponding to site specific $g$ factors \cite{komeda,sainoo} as well as to observation of hyperfine coupling \cite{review}. We further note that the spatial dependence of the signal shows a non-monotonic contour plot, i.e. the signal is elongated and is maximal at $\sim 1$nm on either side of a minimum point \cite{manassen1,manassen2}.

The theoretical understanding of the ESR-STM effect is not settled \cite{review}. The emergence of a finite frequency in a steady state stationary situation is a non-trivial phenomenon. An obvious mechanism for coupling the charge current to the spin precession is spin-orbit coupling \cite{mozyrsky}. It was shown that an ESR signal is present in the noise with spin-orbit coupling when the leads are polarized, either for a strong Coulomb interaction \cite{bulaevskii,gurvitz,martinec} or for the non-interacting case \cite{gurvitz}, and even in linear response \cite{entin}. However, the experimental data \cite{review} involves a small field of $\sim 200$G corresponding to a Larmor frequency of $\sim 500$MHz, i.e. $\sim 10^{-7}$ relative to a lead's bandwidth.
It was found in these spin-orbit models \cite{bulaevskii,gurvitz,martinec} that the signal vanishes when the lead polarization vanishes, or when the lead and dot polarization are parallel, as for a uniform magnetic field.
It was argued that an effective spin polarization is realized as a fluctuation effect either for a small number of electrons that pass the localized spin in one cycle \cite{balatsky} or due to 1/f magnetic noise of the tunneling current \cite{manassen5}.
It was further shown that spin-orbit coupling in an asymmetric dot can yield an oscillating electric dipole, possibly affecting the STM current \cite{levitov}.

 In the present work we follow a recently proposed model that allows for an ESR-STM phenomena with non-polarized leads \cite{horovitz}. The model assumes an additional direct tunneling between the tip and the substrate in parallel to tunneling via the dot's states, i.e. a nanoscopic interferometer. The numerical study \cite{horovitz} shows that the interference of the direct current and that via the spin has an ESR signal in the noise, a signal that increases with the direct tunneling.
  This model is motivated by studies of quantum dots with spin-orbit \cite{lopez} and by STM studies of a two-impurity Kondo system that shows a significant direct coupling between the tip and substrate states \cite{bork}. Similar models including a Aharonov-Bohm phase have been studied \cite{hof,rosa,golub,konig}.
  The nanoscopic interferometer model is consistent with the unusual non-monotonic contour plot \cite{manassen1,manassen2}, i.e. the signal is maximized when the STM tip is not directly on the spin center but slightly away, so as to maximize an overlap with a surface state of the substrate. In the present work we consider non-polarized leads, as relevant to the experimental setup, and evaluate the noise analytically in the stationary system, in accord with the numerical results for this case. The analytic results clarify the physical processes of the resonance phenomenon and allow us to discuss the ESR-STM effect for a broad range of parameters, as in the conclusion section below.

The paper is organized as follows. In Sec.II we introduce the Hamiltonian of the system and present the results for direct tunneling: effective action, the current and the current noise power spectrum.
Sec. III contains the effective action of the dot and the expression for the current flow through dot.
 Sec. IV reflects  our principal result: the resonance part of the current spectral density. The results are illustrated by Figs 1,2. Finally our conclusions are contained in Sec.V. The appendices A,B,C give various details of the calculations.

\section{Hamiltonian}
The Hamiltonian of the system  describes direct tunneling through the dot  between left (L) and right (R) leads as well as L-R tunneling via the dot states,
\begin{equation}\label{Htot}
H=H_L +H_R+H_D+H_{W}+H_T
\end{equation}
 where the lead Hamiltonians are $H_l=\sum_{l,k,\sigma}\epsilon_{l,k}c^{\dagger}_{l,k,\sigma}c_{l,k,\sigma}$, $l=L,R$, $\sigma=\pm$ is the spin and $k$ are the continuum states. The dot Hamiltonian is $H_D=\sum_{\sigma}\epsilon_{\sigma}d_{\sigma}^{\dagger}d_{\sigma}$ with $\epsilon_{\sigma}=\epsilon_0+\sigma H$,  $\epsilon_0$ is the mean position of the dot  levels and $H$ is the applied magnetic field that includes the $g$ factor and the Bohr magneton.
We assume that the dispersions $\epsilon_{l,k}$ of the lead electrons are spin independent, justified by the small ratio $\sim 10^{-7}$ of the Larmor frequency and a typical electron bandwidth.

 A general spin-orbit coupling involves unitary matrices that can be parameterized \cite{horovitz} by two angles $\phi,\theta$. The angle $\phi$ appears in the  Hamiltonian for the direct tunneling
  \begin{equation}\label{HLR}
   H_{W}=W \sum_{k,\sigma}\eexp{i\sigma\phi} c^{\dagger}_{L,k,\sigma}c_{R,k,\sigma}+H.C.
  \end{equation}
The spin dependent form in $\eexp{i\sigma\phi}$ is required by time reversal. The angle $\theta$ appears in the tunneling via the dot as a spin rotation in the $R$ lead, while the $L$ lead is diagonal in spin
  \begin{equation}\label{HT}
   H_T=t\sum_{k,\sigma}[ c^{\dagger}_{L,k,\sigma}d_{\sigma}+c^{\dagger}_{R,k,\sigma}U_{\sigma,\sigma'}d_{\sigma'}]+H.C.
  \end{equation}
 where
 \begin{equation}\label{U}
U=\left(
    \begin{array}{cc}
      \cos(\theta/2) & \sin(\theta/2) \\
      \sin(-\theta/2) & \cos(\theta/2) \\
    \end{array}
  \right)
\end{equation}
We note that special cases of this parameterization have been used in related models \cite{gurvitz,hof,rosa}.

  To calculate the current and the noise
 we use the Keldysh formalism \cite{lifshitz,kamenev} and include in the action a quantum source field $\hat{\alpha} $ that couples to the total current. The source term has the form $\hat{\alpha} =\frac{1}{2}\alpha\sigma_x$ where $\sigma_x$ is a Pauli matrix in the rotated Keldysh space.  The total action is $ S_{tot}= S_T+S_{D}+S_{W} $ where $S_T$  corresponds to $H_T$, i.e. tunneling via the dot
 \begin{eqnarray}\label{ST}
  S_T&=& -t\int dt\{ c_{L,\sigma}^{\dagger}(0)(1-\hat{\alpha}/2)d_{\sigma}+\nonumber\\
  && c^{\dagger}_{R,\sigma}(0)(1+\hat{\alpha}/2)U_{\sigma,\sigma'}d_{\sigma'}+H.C.\}
  \end{eqnarray}
  and $S_D$ is the action of the dot
  \begin{equation}
   S_{D} = \int dt d^{\dagger} G^{-1}_{0}d \label{SD0}\,.
                \end{equation}
$G_0$ is the Green's function (GF) of  the noninteracting dot  in the rotated Keldysh representation that has the form
  \begin{equation}\label{G0}
G_0=\left(
\begin{array}{cc}
G^R_0 & G^K_0 \\
0& G^A_0 \\ \end{array} \right)
  \end{equation}
  with retarded ($R$), advanced ($A$) and Keldysh ($K$) indices as superscripts.
The current via the dot $\delta S_T/\delta\alpha$ is chosen as a symmetric combination of the current from the left lead to the dot $J_{L\rightarrow d}$ and that from the dot to the right lead $J_{d\rightarrow R}$. In general a linear combination of these currents is needed, depending on various capacitances \cite{buttiker}. We expect that the resonance effect is dominated by single occupancy of the dot and the latter two currents are equal.
Indeed we check below that our results for the resonance term do not depend on which linear combination is used. Furthermore, the numerical study \cite{horovitz} used $J_{L\rightarrow d}$ for the noise evaluation with results consistent with our analytic ones.

Here and below the dot electron operator $d$ becomes a vector in spin space (and Keldysh space). The GF $G^{R,A,K}_0$ are diagonal in spin space and in terms of a Fourier energy variable $\epsilon$ are given by
\begin{eqnarray}\label{G0sigma}
G^R_{0,\sigma}(\epsilon)&=&\frac{1}{\epsilon-\epsilon_0 -\sigma H+i\delta}\nonumber\\
G^A_{0,\sigma}(\epsilon)&=&\frac{1}{\epsilon-\epsilon_0 -\sigma H-i\delta}\nonumber\\
G^K_{0,\sigma}(\epsilon)&=&-2\pi i \tanh(\frac{\epsilon}{2T})\delta(\epsilon-\epsilon_0-\sigma H)
\end{eqnarray}
with the limit $\delta=+0$.

 The part of the action $S_{W}$ which contains the leads and the direct LR tunneling is
\begin{eqnarray}
  S_{W} &=& \int dt\sum_{k,k'}c_{k\sigma}^{\dagger} g^{-1}_{kk'\sigma}c_{k'\sigma} \nonumber\\
  g^{-1}_{kk'\sigma} &=& g^{-1}_{k\sigma}\delta_{kk'}-W[e^{i\phi\sigma}A_{kk'}\rho_+(1+\hat{\alpha})+\nonumber\\
 && e^{-i\phi\sigma}A_{kk'}\rho_-(1-\hat{\alpha})]\,.\label{SW}
\end{eqnarray}
 Here $c_{k\sigma}$,  $\rho_{\pm}=(\rho_x\pm i\rho_y)/2$ are  vectors and Pauli matrices, respectively, in LR (left-right) space, the  GFs of the leads $g^{-1}_{k\sigma}$ are diagonal in LR space and
$A_{kk'}=1$ present a constant matrix in momentum k,k' space. Fermion operators and GF as well the quantum source field $ \hat{\alpha}=\alpha \sigma_x $ acts in the rotated Keldysh space. The voltage $V$ between the leads is assumed small relative to the bandwidths, hence the density of states $N_R,N_L$ are taken as constants.
The GF $g_{k\sigma}$ has the structure of Eq. (\ref{G0}) and its momentum integrated forms $\bar{g}_l^{R,A,K}=\sum_k g_{l,k}^{R,A,K}/(2\pi N_l)$ are
\begin{eqnarray}\label{g}
{\bar g}^{R}&=&\frac{1}{2\pi N_l}\sum_k\frac{1}{\epsilon-\epsilon_{l,k}+i\delta}=-\half i \nonumber\\
{\bar g}^{A}&=&\frac{1}{2\pi N_l}\sum_k\frac{1}{\epsilon-\epsilon_{l,k}-i\delta}=+\half i
\nonumber\\
{\bar g}_{R,L}^{K}(\epsilon)&=&-if_{R,L}(\epsilon)
\end{eqnarray}
where $V$ is the voltage difference between the LR leads and $f_{R,L}(\epsilon)= \tanh\frac{\epsilon\mp V/2}{2T}$.

To cope with scattering of electrons due to the tunneling  we shift the operators $c_{k\sigma}$ so as to cancel the linear coupling to $d_\sigma$ in Eq. (\ref{ST}). This adds a term of the form $d^\dagger Q(\alpha)d$ to the dot action and then
the effective action separates into two independent parts $S_{tot}=S_W+S_{dot}$. The total generating functional $Z_{tot}(\alpha)$, as a function of the source field, is therefore factorized into $Z_{tot}=Z_{W} Z_{dot}$.

We consider now the $S_{W}$ part of the effective action.
Inverting $g^{-1}$ by using blockwise matrix inversion  we obtain
\begin{eqnarray}
  g_{LLkk'} &=& g_{Lk} \delta_{kk'} +g_{Lk}x_L D^{-1}_L g_{Lk'}\nonumber\\
  g_{RRkk'} &=& g_{Rk} \delta_{kk'} +g_{Rk}x_R D^{-1}_R g_{Rk'}\label{LL}
\end{eqnarray}
Here $D_{L,R}=1-4x\bar{g}_{L,R} (1\pm\hat{\alpha})\bar{g}_{R,L}(1\mp \hat{\alpha})$, $x_{L,R}=2\pi N_{R,L}W^2(1\pm\hat{\alpha})\bar{g}_{R,L}(1\mp \hat{\alpha})$ and the coupling parameter
\begin{equation}
x=\pi^2N_L N_R W^2
\end{equation}
 The electron transport and noise calculations involve only the integrated GF of Eq. (\ref{g}).

 Direct integration over electron operators $c_{l,k,{\sigma}}$ yields $Z_{W}=det [g^{-1}]=\exp[\Tr\ln g^{-1}]$.
The direct current $J_W$ and related noise power are defined as derivatives with respect to the source field (taking $\alpha=0$ after derivatives is implied)
\begin{eqnarray}
  J_W(t) &=& \frac{\delta \ln Z_W} {\delta \alpha(t)}=\frac{\delta \Tr\ln(g^{-1})}{\delta \alpha(t)}\label{DZ}\\
   S_W  &=& \frac{\delta^2 \ln Z_W}{\delta \alpha(t)\delta \alpha(t')} \label{DDZ}
\end{eqnarray}
We obtain the textbook results \cite{buttiker} for noise and  transport current through a contact with transmission probability \cite{hof}
 $T_B=4x/(1+x)^2$  and reflection coefficient $R_B=1-T_B$ (details in Appendix A), the conductance is then $\frac{2e^2}{h}T_B$.

\section{ Effective Action}
The effective action of the dot includes the $Q(\alpha)$ term from the integration over the lead fermions. It is expressed (in Keldysh space) in terms of various GFs ${\bar g}_{ll'}$ as listed in  Eq. (\ref{gbar}) and in terms of the noninteracting dot GF Eq. (\ref{G0}) as
\begin{eqnarray}
  S_{dot} &=& \int dt d^{\dagger} G^{-1}d \label{SD}\\
  G^{-1} &=& G^{-1}_0-Q(\alpha)\nonumber\\
 Q(\alpha)&=& \Gamma_L (1-\frac{\hat{\alpha}}{2})\hat{g}_{LL}(1+\frac{\hat{\alpha}}{2})+ \Gamma_R (1+\frac{\hat{\alpha}}{2})\hat{g}_{RR}(1-\frac{\hat{\alpha}}{2})+\nonumber\\
 &&2\sqrt{x \Gamma_L \Gamma_R}[ (1+\frac{\hat{\alpha}}{2})M^{\dagger}\hat{g}_{RL}(1+\frac{\hat{\alpha}}{2})+\nonumber\\
 && (1-\frac{\hat{\alpha}}{2})\hat{g}_{LR}M(1-\frac{\hat{\alpha}}{2})]\label{Q}
\end{eqnarray}
 The matrix $M$ is  $M=e^{i\phi\tau_z}U $=$I\nu +i \vec{n}\vec{\tau}$ where $\nu=\cos\theta \cos \phi $, $\vec{n}=(\sin\phi\sin\theta/2, \cos\phi\sin\theta/2, \sin\phi\cos\theta/2)$ and $\tau_i$ are Pauli matrices in spin space. Here we introduce the tunneling widths: $\Gamma_{L,R}=2\pi N_{L,R}t^2$. Taking $\alpha=0$ and
inverting $G^{-1}$ we find the GFs of the dot interacting with the leads (see Appendix A).
As we find below, the resonance contribution to the noise is related to terms that involve the matrices $M,M^\dagger$ (or $C(\epsilon,\epsilon')$ in Eq. (\ref{C5})). We note therefore that the result for the resonance term does not depend on which combination of currents $J_{L\rightarrow d}$ and $J_{d\rightarrow R}$ (determining the source terms $\alpha$ in (\ref{Q})) are used.

Integrating out the dot fermions $d$ with the action (\ref{SD}) we arrive at the generating functional $Z_{dot}(\alpha)=\det G^{-1}=\exp[\Tr\ln G^{-1}]$ which depends on the vertex function $Q(\alpha)$. Similar to (\ref{DZ}) the current through the dot is
\begin{eqnarray}
  J_d(t) &=& \frac{\delta \ln Z_d}{\delta \alpha(t)}=-\Tr[G \frac{\delta Q(\alpha)}{\delta \alpha(t)}]\label{ID}
\end{eqnarray}
Performing calculations for the case of equal tunneling widths $\Gamma_L=\Gamma_R$ (see Appendix B) we obtain the current: $J_{dot}=e(J_{d1}+J_{d2}+J_{d3})$ where
\begin{eqnarray}
  J_{d1} &=& \bar{\Gamma}(1-2T_B)\sum_{\sigma} \int \frac{d\epsilon}{2\pi}ImG^R_{\sigma}(\epsilon)\Delta^{(-)}_{\epsilon} \label{ID1}\\
  J_{d2} &=& -2\nu\bar{\Gamma}\sqrt{R_BT_B}\sum_{\sigma} \int \frac{d\epsilon}{2\pi}ReG^R_{\sigma}(\epsilon)\Delta^{(-)}_{\epsilon}\label{ID2} \\
  J_{d3} &=& T_B\bar{\Gamma}^2 \int \frac{d\epsilon}{2\pi}\Tr[G^R(\epsilon)(\vec{n}\vec{\tau})G^A(\epsilon)(\vec{n}\vec{\tau})]
  \Delta^{(-)}_{\epsilon}\label{JD3}
\end{eqnarray}
here $\bar{\Gamma}=(\Gamma_L+\Gamma_R)/(2(1+x))$ and  $\Delta^{(\mp)}_{\epsilon}=f_L(\epsilon)\mp f_R(\epsilon)$.
 The trace in the last equation (\ref{JD3}) is
\begin{eqnarray}
  \Tr[...] &=& \sum_{\sigma}[-\frac{\sin^2\phi \cos^2\theta/2}{\bar{\Gamma}}
  ImG^R_{\sigma}(\epsilon)+\nonumber\\
  && \sin^2(\theta/2) G^R_{\sigma} G^A_{-\sigma}]\label{tr}
\end{eqnarray}
 presents two different scattering processes. The first term with $\sin^2\phi$ appears also  for  an Aharonov-Bohm phase and for $\theta=0$  the corresponding current  coincides with that in Ref. \onlinecite{hof}; this term does not depend on the  sign of phase $\phi$  meantaining the relation $G(\phi)=G(-\phi)$ for conductance in closed system (two-terminal setup).

  We note that for $\theta=0$ the two spin states decouple and a resonance phenomena at the Larmor frequency is not possible. There are still interference effects due to the phase $\phi$, though these are unrelated to the resonance.
 The second term of Eq. (\ref{tr}) describes the spin orbit effect and reflects tunneling transitions accompanied by spin flips. The phase $\theta$ is therefore controlling the ESR effect.

\section { Current Spectral Density}
The current noise power $S_d$ is given by formula (\ref{DDZ}) in which $Z_W$ is replaced by $Z_{dot}$. The total noise function can be written as a sum of two terms $S_d(t,t')=S_{d1}(t,t')+S_{d2}(t,t')$

\begin{eqnarray}
  S_{d1}(t,t') &=& -\Tr[G\frac{\delta^2 Q}{\delta \alpha(t)\delta \alpha(t')}]\\
  S_{d2}(t,t')&=&-\Tr[G\frac{\delta Q}{\delta \alpha(t)}G\frac{\delta Q}{\delta \alpha(t')}]\label{Stt}
\end{eqnarray}
We calculate the current spectral density to lowest order in $W$, these terms have a  resonance contribution at Larmor frequency. Details of the derivation are given in Appendix C. We write the frequency dependent noise $S_{d1}(\omega)$ and $S_{d2}({\omega})$  as an expansion in $W\sim\sqrt{x}$:
\begin{eqnarray}
  S_{d1}(\omega) &=& S^0(\omega)+\sqrt{x}S^1(\omega)+xS^{(2)}(\omega)\nonumber \\
  S_{d2}(\omega) &=& S_0(\omega)+\sqrt{x}S_1(\omega)+xS_2(\omega)\label{STD}
\end{eqnarray}
The spin flip transport which is responsible for the resonance occurs due to the spin-orbit interacting vertices. This effect takes place at least in the terms linear in $x$ ($S^{(2)},S_2$) which are presented below (all other contributions to the noise power are given explicitly in Appendix C). The first term depends weakly on frequency
\begin{eqnarray}
  S^{(2)}(\omega)&=&\frac{3e^2\bar{\Gamma}^2}{4(1+x)^2}\int\frac{d\epsilon}{2\pi}\Tr(G^R(\epsilon)
 \vec{n}\vec{\tau}G^A(\epsilon)\vec{n}\vec{\tau})[\Delta^{(-)}_{\epsilon-\omega}\nonumber\\
&& +\Delta^{(-)}_{\epsilon+\omega}]\Delta^{(-)}_{\epsilon}
\end{eqnarray}
The second term $S_2 $ contains three contributions
\begin{equation}\label{S2}
    S_2(\omega)=S_2^{RR+AA}+S_2^{RA}+S_2^{K}
\end{equation}
where
 \begin{eqnarray}
   S_2^{RR+AA} &=& \frac{e^2\bar{\Gamma}^2}{4(1+x)^2}\int\frac{d\epsilon}{2\pi}(\hat{n}^{RR}+\hat{n}^{AA})
    \Delta^{(+)}_{\epsilon-\omega/2}\Delta^{(+)}_{\epsilon+\omega/2}\nonumber
   \end{eqnarray}
   \begin{eqnarray}
 S_2^{RA} &=& e^2\bar{\Gamma}^2\int\frac{d\epsilon}{2\pi}(\hat{n}^{RA}+\hat{n}^{AR})[1-\nonumber\\
 &&\frac{f_L(\epsilon-\omega/2)f_R(\epsilon+\omega/2)+(\omega\rightarrow -\omega)}{1+x}]\nonumber
 \end{eqnarray}
 \begin{eqnarray}
   S_2^{K} &=& e^2\bar{\Gamma}^2\sqrt{R_B}\int\frac{d\epsilon}{2\pi}\Tr \{\hat{m}
   \Delta^{(+)}_{\epsilon-\omega/2}\Delta^{(+)}_{\epsilon+\omega/2}\nonumber\\
   &&+\frac{\bar{\Gamma}}{1+x}[ImG^R(\epsilon-\omega/2)\vec{n}\vec{\tau}
   G^R(\epsilon+\omega/2)\vec{n}\vec{\tau}\times\nonumber\\
   &&G^A(\epsilon+\omega/2)
   \Delta^{(-)}_{\epsilon-\omega/2}\Delta^{(-)}_{\epsilon+\omega/2}+(\omega\rightarrow-\omega)]\} \nonumber
 \end{eqnarray}
 here
 \begin{eqnarray}
 \hat{n}^{MN}&=&\Tr [G^M(\epsilon-\omega/2)\vec{n}\vec{\tau}
   G^N(\epsilon+\omega/2)\vec{n}\vec{\tau}]\nonumber\\
   \hat{m}&=&\Im G^R(\epsilon-\omega/2)\vec{n}\vec{\tau}
   ImG^R(\epsilon+\omega/2)\vec{n}\vec{\tau}\nonumber
  \end{eqnarray}
where $ M,N=R(A)$.

 The  resonance behavior at $\omega=2H$ that we find is related to $S_2(\omega)$. We note first, as it is easy to check, that the total noise power (at $\omega\rightarrow0$) $S_d\rightarrow 0$ in each order in $W$ if $T>V$ and  $T\rightarrow 0$. This observation serves as additional test for our calculations.

 Separating  the resonance contributions in the expression for $S_2(\omega)$
  yields at the principal result of our work:
 $S_2^{sing}= S_2^{r}+ S_2^{s}$
 where
  \begin{eqnarray}
  S_2^{r} &=& 2e^2\bar{\Gamma}^2\sin^2{\theta/2}\sum_\sigma\int\frac{d\epsilon}{2\pi} ImG^R_{\sigma}(\epsilon-\omega/2)\times\nonumber\\
   &&ImG^R_{-\sigma}(\epsilon+\omega/2)F(\epsilon,\omega);
    \end{eqnarray}
   \begin{eqnarray}
  F(\epsilon,\omega)&=&1-\frac{1}{4}[ 3(f_L(\epsilon-\omega/2)f_R(\epsilon+\omega/2)
  +(\omega\rightarrow -\omega)) \nonumber\\
  && -(f_R(\epsilon-\omega/2)f_R(\epsilon+\omega/2)+(R\rightarrow L))]
    \label{F}
  \end{eqnarray}
    \begin{eqnarray}
   S_2^{s} &=& e^2\bar{\Gamma}^3\sin^2{\theta/2}\sum_\sigma\int\frac{d\epsilon}{2\pi} [ImG^R_{\sigma}(\epsilon-\omega/2)\times\nonumber\\
   &&ImG^R_{-\sigma}(\epsilon+\omega/2)ImG^R_{\sigma}(\epsilon+\omega/2)\nonumber\\
   &&\Delta^{(-)}_{\epsilon-\omega/2}\Delta^{(-)}_{\epsilon+\omega/2}+(\omega\rightarrow-\omega)]\label{sr2}
  \end{eqnarray}
There is a potentially singular contribution that also in $S^{(2)}$
\begin{eqnarray}\label{S2s}
  S^{(2)sing}&=&\frac{3e^2\bar{\Gamma}^2}{4}\sin^2{\theta/2}\sum_\sigma\int\frac{d\epsilon}{2\pi}ImG^R_{\sigma}(\epsilon)
 ImG^R_{-\sigma}(\epsilon)\times \nonumber\\
 &&[\Delta^{(-)}_{\epsilon-\omega}
 +\Delta^{(-)}_{\epsilon+\omega}]\Delta^{(-)}_{\epsilon}
\end{eqnarray}
 However, it can be seen that this term depends weakly on frequency
and therefore it does not have a resonance at the Larmor frequency.

 In the experiments \cite{review} the parameters satisfy $V>>T>>2H$. Therefore, if the mean level position $\epsilon_0$ is in between the two chemical potentials $-\half V<\epsilon_0<\half V$ and is not too close to $\pm \half V$, i.e. $|\epsilon_0\pm \half V|\gtrsim H$, then we have $F(\epsilon,\omega)\simeq3$ and $\Delta_{\epsilon\pm\omega/2}\simeq2$. We show in particular the function $F(\epsilon_0)$ of Eq. (\ref{F}), neglecting $\omega$ and $H$ terms, in Fig. 1.

 \begin{figure} [!ht]
\centering
\includegraphics[scale=0.7]{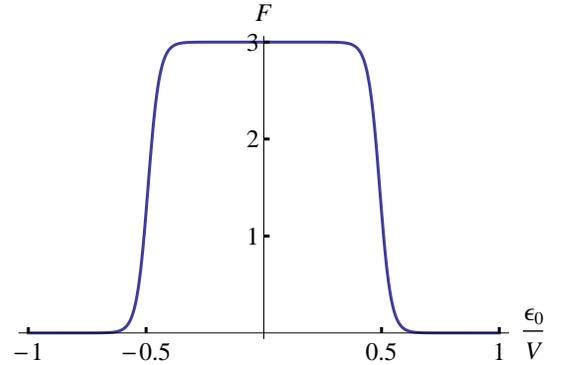}
\caption {Dependence of the noise on the mean level position $\epsilon_0$, the function $F(\epsilon_0)$ in Eq. (\ref{F}) neglecting $\omega,H$ terms (i.e. valid for $|\epsilon_0\pm \half V|\gtrsim H$). The voltage and temperature ratio is $\frac{V}{T}=40$.}
\end{figure}
 Taking $F(\epsilon,\omega)\simeq3$ and $\Delta^-_{\epsilon\pm\omega/2}\simeq2$ and using Eq. (\ref{imG}) the integrals are simply evaluated
\begin{eqnarray}
  S_2^{r} &\simeq & \frac{3\pi e^2\sin^2\half\theta}{2\pi}\sum_\sigma\frac{\bar{\Gamma}^3}{(\omega/2+\sigma H)^2+\bar{\Gamma}^2}\label{sing1}
  \end{eqnarray}
  \begin{eqnarray}
  S_2^{s} &\simeq &\frac{ \pi e^2\bar{\Gamma}^5\sin^2\half\theta}{2\pi(H^2+\bar{\Gamma}^2)}\sum_\sigma\frac{[(\half\omega+\sigma H)^2+3\bar{\Gamma}^2-\half\sigma\omega H]}{[(\half\omega+\sigma H)^2+\bar{\Gamma}^2](\bar{\Gamma}^2+\omega^2)}\nonumber\\\label{sing2}
\end{eqnarray}
In standard units the $2\pi$ in the denominators are replaced by $h$. Thus the last equations (\ref{sing1})
and (\ref{sing2}) show the  resonance behavior of the noise power at $\omega=\pm2 H$.

\begin{figure} [!ht]
\centering
\includegraphics [width=0.45 \textwidth ]{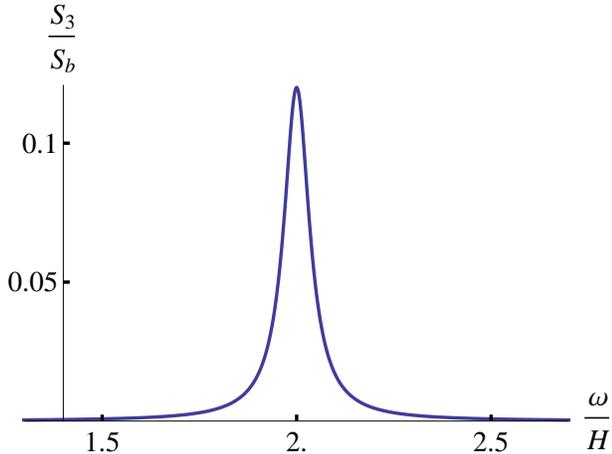}
\caption {The  current spectral  power $S_3(\omega)$ which is a sum of two resonance contributions (\ref{sing1}) and (\ref{sing2}) showing resonance peaks at  $\omega=2 H$. The width parameter is $\bar{\Gamma}/H= 0.02$. }
\end{figure}

In Fig. 2 we plot the  noise power $S_3(\omega)= S_2^{r}+S_2^{s}$, Eqs. (\ref{sing1}) and (\ref{sing2}), normalized by $S_b=\frac{\pi}{2h} e^2H\sin^2\half\theta$  as function of frequency.

For a sharp resonance, as seen experimentally \cite{review} ${\bar\Gamma}\ll 2H$ since ${\bar\Gamma}$ determines the resonance width. Therefore the ratio $S_2^s/S_2^r\approx
{\bar\Gamma}^2/H^2\ll 1$ is small and $S_2^r$ with the Lorenzian shape dominates.  In the wide range where $F=3$ (Fig. 1) we have therefore for the signal amplitude at resonance
\begin{equation}
S_{signal}=\frac{e^2}{h}3\pi {\bar\Gamma}x\sin^2\half\theta
\end{equation}
We note also that at $x\gg 1$ where ${\bar\Gamma}x\rightarrow \Gamma$ another term causes a cancellation of this one (appendix C) and the signal vanishes at large $x$. We expect then that the signal is maximal at $x\approx 1$, in accord with numerical data \cite{horovitz}.

\section{Conclusion} We have considered a general spin-orbit scattering mechanism in a setup of a nanoscopic interferometer and have shown that the interference of the two transmission paths leads
to a resonance contribution to the current correlation spectral density at the Larmor frequency. In particular we find that the effect takes place in the absence of lead polarizations, consistent with ESR-STM experiments. Our model also accounts for several unusual features of the data: (i) A sharp resonance even at high temperatures $T\gg H$, (ii) insensitivity to the details of the spin defect, i.e. to the positions of its levels between the tip and substrate chemical potentials, (iii) contour plots \cite{manassen1,manassen2} showing that the signal is maximal at $\sim 1$nm from a center, hence a significant direct coupling $W$ bypassing the spin can be achieved.

Here we have neglected the Coulomb interaction $U$ between charges on the dot. However, for experimentally interesting case of a large applied voltage $V$,  the Coulomb repulsion is expected to satisfy $U <<eV$. The levels therefore remain between the two chemical potentials and we expect that the resonance part of the noise is weakly affected by $U$. A similar conclusion was reached for the case with polarized leads \cite{gurvitz}. We note also that the insensitivity of the resonance term to the choice of $J_{L\rightarrow d}$ or $J_{d\rightarrow R}$ to represent the current via the dot implies that the charge occupancy of the dot is constant, i.e. singly occupied. Therefore, the Coulomb interaction on the dot is not expected to affect the resonance term.

Our key result Eq. (\ref{sing1}) shows that the signal amplitude is $S_{signal}=\frac{e^2}{h}3\pi\sin^2\half\theta {\bar\Gamma}x \sim t^2W^2$. The signal should vanish at $W=0$ on general grounds \cite{horovitz}, yet the $W^2$ form is unexpected. Some of our other results can be obtained for small $t,W$ by simple estimates: The resonance linewidth follows from a golden rule $\Gamma=2\pi t^2N(0)$; the DC current via the dot $J_d=2e\Gamma/\hbar$ for $eV\gg 2H, T$ (Eq. (\ref{ID1}) is the dominant term) corresponds to a transition rate $\Gamma/\hbar$ from either reservoir to the dot, hence $J_d=2e\Gamma/\hbar$ given the dot's two states. The direct transport of $L\rightarrow R$ is also a golden rule $\Gamma_W=2\pi W^2N_L(0)$ times the final number of states $N_R(0)eV$, hence $J_W=\frac{2e^2}{h} 4xV$ (for $x\ll 1$) while the corresponding background noise is a classical shot noise $S_W=2eJ_W$, Eq. (\ref{SW}). The noise of the dot current is, however, much reduced from that of a shot noise since $S^0\approx \frac{1}{4} eJ_d$, Eq. (\ref{S0}).

To analyze the experimental data we first estimate the relevant parameters. The resonance linewidth is $\sim 1$MHz $=\Gamma/2\pi=t^2N(0)/\hbar$ (for $x\ll 1$). Assuming a metallic $N(0)\sim 1/(5$eV) yields $tN(0)\approx 10^{-5}$. Considering next the DC current $0.1-1$nA at $\sim 1$V : The dot current for $x\ll 1$ $J_d=2e\Gamma/\hbar\approx 10^{-12}$A is too small, hence the DC current is dominated by the direct coupling $W$ with $J_W=\frac{2e^2}{h}4x V$, hence $WN(0)\approx 10^{-3}$ and $W\gg t$. The background noise due to the dot current is $S_0\approx e^2\Gamma/\hbar$ while that from $W$ is $S_W=\frac{2e^2}{h} 8xeV$, hence $\frac{S_0}{S_W}\approx\frac{t^2}{W^2}/[8N(0)eV]\ll 1$, i.e. the background noise is dominated by $S_W$.

We note that the background noise is not measured in the experiment since the modulation technique \cite{review} measures the derivative of the noise spectra. Furthermore, the signal intensity is under study \cite{manassen4} as it is highly sensitive to uncertainties in the feedback and impedance matching circuits. We find that the signal to background intensity for $x\ll 1$ is $S_{signal}/S_W=\frac{3\pi}{16}\sin^2\half\theta\frac{\Gamma}{eV}$, i.e. of order $10^{-9}-10^{-8}$.  In conclusion, our model presents an analytic solution to a long standing puzzle, paving the way for more controlled single spin detection via ESR-STM.

\begin{acknowledgments}
We thank for stimulating discussions with Y. Manassen, O. Entin-Wohlman, S. A. Gurvitz, A. Janossy, L. S. Levitov, I. Martin, M. Y. Simmons, F. Simon, E. I. Rashba, S. Rogge, A. Shnirman, G. Z\'{a}rand and A. Yazdani. This research was supported by THE ISRAEL SCIENCE FOUNDATION (BIKURA) (grant No. 1302/11) and by
the Israel-Taiwanese Scientific Research Cooperation of the Israeli Ministry of Science and Technology.
\end{acknowledgments}

\appendix



\section{Direct Current and Noise. Green's Functions}
The GFs integrated over momentum are obtained by inverting the inverse Green function in Eq. (\ref{SW}), as shown for the diagonal terms in Eq. (\ref{LL}). Here we write the whole list of these functions:
\begin{eqnarray}
 \bar{g}_{LL} &=& 2\pi N_L D^{-1}_L \bar{g}_{L}= 2\pi N_L \hat{g} _{LL}\nonumber\\
  \bar{g}_{RR}  &=& 2\pi N_R D^{-1}_R \bar{g}_{R}= 2\pi N_R \hat{g} _{RR}\nonumber\\
  \bar{g}_{RL}  &=& 2\pi N_R W \bar{g}_{R}(1-\hat{\alpha})e^{-\phi\sigma}\bar{g}_{LL}= \nonumber\\
  &&(2\pi)^2 N_R N_L W e^{-\phi\sigma}\hat{g} _{RL}\nonumber\\
  \bar{g}_{LR}  &=& 2\pi N_R W \bar{g}_{LL}(1+\hat{\alpha})e^{\phi\sigma}\bar{g}_{R}=\nonumber\\
 && (2\pi)^2 N_R N_L W \hat{g}_{LR}e^{\phi\sigma} \label{gbar}
\end{eqnarray}
Explicitly for $\alpha=0$ we can write
\begin{eqnarray}
  \bar{g}_{LL}^{R,A}(\epsilon) &=& 2\pi N_L\frac{\mp i}{2(1+x)} \label{g1}\\
  \bar{g}_{LL}^{K}(\epsilon) &=& \frac{-2\pi N_L i}{(1+x)^2}(f_L(\epsilon)
  +xf_R(\epsilon))\label{g2}
\end{eqnarray}
Changing R to L  yields $\bar{g}_{RR}$.
The off-diagonal functions acquire a form
\begin{eqnarray}
  \hat{g}_{RL}^{R,A} &=& -\frac{1}{4(1+x)},\,\,
  \hat{g}_{RL}^{K} = \frac{-1}{2(1+x)^2}\Delta^{(-)}_{\epsilon} \label{g3} \\
  \hat{g}_{LR}^{R,A} &=& -\frac{1}{4(1+x)},\,\,
  \hat{g}_{LR}^{K} =\frac{1}{2(1+x)^2}\Delta^{(-)}_{\epsilon}\label{g4}\\
  \Delta^{(\pm)}_{\epsilon}&=&\tanh\frac{\epsilon+ V/2}{2T}
   \pm\tanh\frac{\epsilon- V/2}{2T}
\end{eqnarray}
With the help of these functions we find the direct tunneling current and the corresponding noise power which acquire standard forms (below $\sigma_i$ are Pauli matrices that act in Keldysh space)
\begin{eqnarray}
  J_W(t) &=& \Tr[g_{LR}\frac{\delta g_{LR}^{-1}}{\delta \alpha(t)}+g_{RL}\frac{\delta g_{RL}^{-1}}{\delta \alpha(t)}] \\
 J_W &=&2eT_B\int \frac{d\epsilon}{2\pi}\Delta^{(-)}_{\epsilon}
\end{eqnarray}
and
\begin{eqnarray}\label{SW}
   S_W  &=& \Tr[(\frac{\delta g_{LR}(\bar{t}t')}{\delta \alpha(t)}-
   \frac{\delta g_{RL}(\bar{t}t')}
    {\delta\alpha(t)})\sigma_x]  \nonumber \\
 S_W (0)&=& \frac{4e^2}{2\pi}[eV T_B(1-T_B)\coth\frac{V}{2T}+2T_B^2T]\label{SW}
\end{eqnarray}
The noise $S_W(\omega)$ is well known \cite{buttiker} and coincides with Eq. (\ref{SW}) for small $\omega$, $\omega\ll eV$.
The effective action of the  dot is given by Eq. (\ref{SD}). In the limit of vanishing source terms the corresponding GFs are obtained by
inverting $G^{-1}$ ($\alpha=0$)
\begin{eqnarray}
  G^R_{\sigma}(\epsilon) &=& \frac{1}{\epsilon-\epsilon_{\sigma}+r+\frac{i(\Gamma_L +\Gamma_R)}{2(1+x)}} \label{g5}\\
   r &=& \frac{\nu\sqrt{x\Gamma_L\Gamma_R}}{1+x}\nonumber\\
   G^R_{\sigma}(\epsilon)G^A_{\sigma}(\epsilon) &=&- \frac{2(1+x)}{\Gamma_L+\Gamma_R}ImG^R_{\sigma}(\epsilon) \\
   ImG^R_{\sigma}(\epsilon)&=&\frac{-\bar{\Gamma}}{(\epsilon-\epsilon_{\sigma}+r)^2+
   \bar{\Gamma}^2},\,\,\,\label{imG}
  \end{eqnarray}
   \begin{eqnarray}
  G^K(\epsilon) &=&\frac{2iImG^R(\epsilon)}{(1+x)(\Gamma_L+\Gamma_R)}[f_L(\epsilon)
  (\Gamma_L+x\Gamma_R)+\nonumber\\
  &&f_R(\epsilon)(x\Gamma_L+\Gamma_R)]+ \nonumber\\
  &&2i\frac{\sqrt{x \Gamma_L\Gamma_R}}{(1+x)^2}\Delta^{(-)}_{\epsilon}G^R(\epsilon)\vec{n}
  \vec{\tau}G^A(\epsilon)\label{g6}
\end{eqnarray}

\section {Current through the dot}
 Next we calculate the transmission through the dot which is presented by Eq. (\ref{ID}).
  The superscripts in the following correspond to matrix elements in Keldysh space,
\begin{eqnarray}
  J_d(t) &=& \Tr[G^R (\frac{\delta Q(\alpha)}{\delta \alpha(t)})^{11}+G^K (\frac{\delta Q(\alpha)}{\delta \alpha(t)})^{21}\nonumber\\
  &&+G^A (\frac{\delta Q(\alpha)}{\delta \alpha(t)})^{22}]\nonumber\\
  \frac{\delta Q(\alpha)}{\delta \alpha(t)} &=& \frac{1}{4}[-\Gamma_L(\sigma_x \hat{g}_{LL}(t\bar{t})(1+\frac{\hat{\alpha}}{2})\nonumber\\
  &&-(1-\frac{\hat{\alpha}}{2})\hat{g}_{LL}(\bar{t}t)\sigma_x)+\nonumber\\
 && \Gamma_R(\sigma_x \hat{g}_{RR}(t\bar{t})(1-\frac{\hat{\alpha}}{2})-(1+\frac{\hat{\alpha}}{2})
  \hat{g}_{RR}(\bar{t}t)\sigma_x)+\nonumber\\
  &&2\sqrt{x \Gamma_L \Gamma_R} (\sigma_x M^{\dagger}\hat{g}_{RL}(1+\frac{\hat{\alpha}}{2})+\nonumber\\
  &&(1+\frac{\hat{\alpha}}{2}) M^{\dagger}\hat{g}_{RL}\sigma_x-\nonumber\\
  &&\sigma_x \hat{g}_{LR}M(1-\frac{\hat{\alpha}}{2})- (1-\frac{\hat{\alpha}}{2})\hat{g}_{LR}M\sigma_x)]+\nonumber\\
  &&\Gamma_L\frac{\delta \hat{g}_{LL}(t_1t_2)}{\delta \alpha(t)}+\Gamma_R\frac{\delta \hat{g}_{RR}(t_1t_2)}{\delta \alpha(t)}+\nonumber\\
  &&2\sqrt{x \Gamma_L \Gamma_R}[\frac{\delta \hat{g}_{LR}(t_1t_2)}{\delta \alpha(t)}M+M^{\dagger}\frac{\delta \hat{g}_{RL}(t_1t_2)}{\delta \alpha(t)}]\nonumber\\ \label{dQ}
\end{eqnarray}
The current takes the form
\begin{eqnarray}
J_d &=& e\int \frac{d\epsilon}{2\pi} \Tr\{G(\epsilon)\{\frac{1}{2}[ -\Gamma_L(\sigma_x \hat{g}_{LL}(\epsilon)-\hat{g}_{LL}(\epsilon)
  \sigma_x)+\nonumber\\
  &&\Gamma_R(\sigma_x \hat{g}_{RR}(\epsilon)-
  \hat{g}_{RR}(\epsilon)\sigma_x)\nonumber\\
 &&+2\sqrt{x \Gamma_L \Gamma_R} (\sigma_x M^{\dagger}\hat{g}_{RL}(\epsilon)+\nonumber\\
  && M^{\dagger}\hat{g}_{RL}(\epsilon)\sigma_x-
  \sigma_x \hat{g}_{LR}(\epsilon)M- \hat{g}_{LR}(\epsilon)M\sigma_x)]+\nonumber\\
  &&\Gamma_L \delta \hat{g}_{LL}(\epsilon,\omega)+\Gamma_R\delta \hat{g}_{RR}(\epsilon,\omega)+\nonumber\\
  &&2\sqrt{x \Gamma_L \Gamma_R}(\delta \hat{g}_{LR}(\epsilon,\omega)M+M^{\dagger}\delta \hat{g}_{RL}(\epsilon,\omega)\}\}\nonumber
\end{eqnarray}
The variations of the GFs are given as a Fourier transform,
\begin{eqnarray}
\delta \hat{g}_{LL}(\epsilon,\omega) &=& 4x \hat{g}_{LL}(\epsilon-\omega)[\sigma_x \bar{g}_R (\epsilon)-  \nonumber\\
&&\bar{g}_R (\epsilon-\omega)\sigma_x] \hat{g}_{LL}(\epsilon) \nonumber\\
\delta \hat{g}_{RR}(\epsilon,\omega) &=& -4x \hat{g}_{RR}(\epsilon-\omega)[\sigma_x \bar{g}_L (\epsilon)-  \nonumber\\
&&\bar{g}_L (\epsilon-\omega)\sigma_x] \hat{g}_{RR}(\epsilon)\nonumber\\
\delta \hat{g}_{LR}(\epsilon,\omega) &=& \hat{g}_{LL}(\epsilon-\omega)\sigma_x\bar{g}_R(\epsilon)+\nonumber\\
&&\delta g_{LL}(\epsilon,\omega)(1+\hat{\alpha})\bar{g}_R (\epsilon)\nonumber\\
\delta \hat{g}_{RL}(\epsilon,\omega) &=& - \bar{g}_R (\epsilon-\omega)\sigma_x\hat{g}_{LL}(\epsilon) +\nonumber\\
&&\bar{g}_R (\epsilon-\omega)(1-\hat{\alpha})\delta g_{LL}(\epsilon,\omega)\nonumber
 \end{eqnarray}
Performing the trace in Keldysh space and using the explicit form of the lead GFs Eqs. (\ref{g1}-\ref{g4}) as well the dot GFs Eqs. (\ref{g5}-\ref{g6}) we arrive at Eqs. (\ref{ID1}-\ref{JD3}).

\section {Current noise power}
We consider the current noise power for equal tunneling widths $\Gamma_L=\Gamma_R$ to  order $x$. At first we present the derivation of $S_{d1}$. This part of the noise power depends on the second variation of the vertex function $\delta^2 Q$. Their Fourier transformed Keldysh components acquire a form
\begin{eqnarray}
  (\delta^2 Q)^{11}_{\omega}&=&-\frac{\bar{\Gamma}}{4}[i+4\sqrt{x}\nu F_{1\omega}+\frac{i\sqrt{x}\vec{n}\vec{\tau}}{1+x}F_{2\omega}-\nonumber\\
  &&\frac{2ix}{1+x}(4-F_3+F_4)]\nonumber \\
  (\delta^2Q)^{22}_{\omega}&=&\frac{\bar{\Gamma}}{4}[i-4\sqrt{x}\nu F_{1\omega}+\frac{i\sqrt{x}\vec{n}\vec{\tau}}{1+x}F_{2\omega}-\nonumber\\
  &&\frac{2ix}{1+x}(4-F_3+F_4)]\nonumber\\
  (\delta^2 Q)^{21}_{\omega}&=&\frac{i\bar{\Gamma}}{16}[(\Delta^{(+)}_{\epsilon-\omega}+
  \Delta^{(+)}_{\epsilon+\omega})+
  6\frac{\sqrt{x}\vec{n}\vec{\tau}}{1+x}(\Delta^{(-)}_{\epsilon-\omega}+
  \Delta^{(-)}_{\epsilon+\omega})] \label{DDS1}\nonumber
\end{eqnarray}
where
\begin{eqnarray}
  F_{1\omega} &=& 1-\frac{(\Delta_L+x\Delta_R)f_R(\epsilon)+\Delta_R(f_L(\epsilon)+x f_R(\epsilon))}{4(1+x)}\nonumber \\
 F_{2\omega}&=&\Delta_Lf_R(\epsilon)-\Delta_Rf_L(\epsilon)\nonumber\\
 F_3&=&\frac{1}{1+x}\{[(f_L(\epsilon)+x f_R(\epsilon))f_R(\epsilon-\omega)+\nonumber\\
 &&(L \Leftrightarrow R)]+(\omega\rightarrow-\omega)\}\nonumber\\
 F_4&=& \sqrt{R_B}\Delta^{(-)}_{\epsilon}(\Delta^{(-)}_{\epsilon+\omega}+\Delta^{(-)}_{\epsilon-\omega})
\end{eqnarray}
here $\Delta_{L,R} = f_{L,R}(\epsilon-\omega)+f_{L,R}(\epsilon+\omega)$. After tracing Keldysh space we obtain {\color{red} }
\begin{equation}\label{sd1}
    S_{d1} = -\frac{e^2}{2}\Tr[G^R(\delta^2 Q)^{11}+G^K(\delta^2 Q)^{21}+G^A(\delta^2 Q)^{22}]
\end{equation}
Using the explicit forms for vertices (\ref{DDS1}) (see also first Eq. (\ref{STD})) we arrive at
\begin{eqnarray}
 S^{0}(\omega) &=& -e^2\frac{\bar{\Gamma}}{2}\int\frac{d\epsilon}{2\pi}\Tr\{ImG^R(\epsilon)[1-\nonumber\\
 &&\frac{1}{8}
 (f_L(\epsilon)+f_R(\epsilon))(\Delta^{(+)}_{\epsilon-\omega}+
  \Delta^{(+)}_{\epsilon+\omega})]\}\label{S0}
  \end{eqnarray}
 \begin{eqnarray}
 S^1(\omega)&=&-e^2\frac{\bar{\Gamma}}{2}\int\frac{d\epsilon}{2\pi}\Tr\{ImG^R(\epsilon)[\frac{\vec{n}\vec{\tau}}{1+x}
(F_{2\omega}-\nonumber\\
&&\frac{3}{4}(f_L(\epsilon)+f_R(\epsilon))
 (\Delta^{(-)}_{\epsilon-\omega}+
  \Delta^{(-)}_{\epsilon+\omega}))]-\nonumber\\
 &&4\nu F_{1\omega}ReG^R(\epsilon)- \frac{\bar{\Gamma}}{4}G^R(\epsilon)\vec{n}\vec{\tau}G^A(\epsilon)\times\nonumber\\
&&\Delta^{(-)}_{\epsilon}(\Delta^{(+)}_{\epsilon-\omega}+
  \Delta^{(+)}_{\epsilon+\omega})\}\label{SK}
\end{eqnarray}
and the formula for $S^{(2)}$ is given in the main text Eq. (\ref{S2}).

The other part of the current spectral density $S_{d2}(\omega)$ (see Eq.(\ref{Stt})) is defined by Fourier transformed GFs and vertices
\begin{eqnarray}\label{C5}
 S_{d2}(\omega)&=&-e^2\int\frac{d\epsilon}{2\pi} \Tr[G(\epsilon-\frac{\omega}{2})\delta Q(\omega)G(\epsilon+\frac{\omega}{2})\delta Q(-\omega)]\nonumber\\
  2\delta Q(\omega)&=&A(\epsilon+\frac{\omega}{2})+B(\epsilon-\frac{\omega}{2})+\nonumber\\
 && C(\epsilon-\frac{\omega}{2},\epsilon+\frac{\omega}{2})+
  4xD(\epsilon-\frac{\omega}{2},\epsilon+\frac{\omega}{2})+\nonumber\\
 && 8x\sqrt{x}E(\epsilon-\frac{\omega}{2},\epsilon+\frac{\omega}{2})
\end{eqnarray}
where
\begin{eqnarray}
 A(\epsilon)&=&\frac{\Gamma}{2}\sigma_x\{-(\hat{g}_{LL}(\epsilon)-  \hat{g}_{RR}(\epsilon))+\nonumber\\
 && 2\sqrt{x} [ M^{\dagger}g_R(\epsilon)\hat{g}_{LL}(\epsilon)-
  \hat{g}_{LL}(\epsilon)g_R(\epsilon)M]\}\nonumber\\
   B(\epsilon)&=&\frac{\Gamma}{2}\{\hat{g}_{LL}(\epsilon)-  \hat{g}_{RR}(\epsilon)+\nonumber\\
  && 2\sqrt{x} [ M^{\dagger}g_R(\epsilon)\hat{g}_{LL}(\epsilon)-
  \hat{g}_{LL}(\epsilon)g_R(\epsilon)M]\}\sigma_x\nonumber\\
  C(\epsilon,\epsilon')&=&2\sqrt{x} \Gamma[\hat{g}_{LL}(\epsilon)\sigma_x g_R(\epsilon')M-M^{\dagger}g_R(\epsilon)\sigma_x\hat{g}_{LL}(\epsilon')]\nonumber\\
 D(\epsilon,\epsilon')&=& \Gamma[\hat{g}_{LL}(\epsilon)(\sigma_x g_R(\epsilon')-g_R(\epsilon)\sigma_x)\hat{g}_{LL}(\epsilon')]\nonumber\\
 &&-(L\Leftrightarrow R)\nonumber
 \end{eqnarray}
 and
 \begin{eqnarray}
 E(\epsilon-\frac{\omega}{2},\epsilon+\frac{\omega}{2})&=&\hat{Y}g_R(\epsilon+\frac{\omega}{2})M+M^+ g_R(\epsilon-\frac{\omega}{2})\hat{Y}\nonumber
 \end{eqnarray}
 where
 \begin{eqnarray}
 \hat{Y}&=&\Gamma \hat{g}_{LL}(\epsilon-\frac{\omega}{2})[\sigma_x g_R(\epsilon+\frac{\omega}{2})- g_R(\epsilon-\frac{\omega}{2})\sigma_x]\hat{g}_{LL}(\epsilon+\frac{\omega}{2})\nonumber
 \end{eqnarray}
Indeed the vertex function $ D(\epsilon,\epsilon')$ is irrelevant for spin flip processes and may be ignored.
Explicit form for Keldysh components of $\delta Q(\omega)$ to linear order in $x$ can be simply find
\begin{eqnarray}
\delta Q^{21} (\omega)&=& i\bar{\Gamma}\sqrt{x} \vec{n}\vec{\tau}\nonumber\\
\delta Q^{11}(\omega) &=&-\frac{\bar{\Gamma}\sqrt{R_B}}{4}i\Delta^{(-)}_{\epsilon-\omega/2}-
\frac{\bar{\Gamma}\sqrt{x}}{2(1+x)}[(2\nu\Delta^{(-)}_{\epsilon-\omega/2}+\nonumber\\
&&2i\vec{n}\vec{\tau}(xf_R(\epsilon-\omega/2)+\frac{1}{2}\Delta^{(+)}_{\epsilon-\omega/2})]\nonumber\\
\delta Q^{22}(\omega) &=&\frac{\bar{\Gamma}\sqrt{R_B}}{4}i\Delta^{(-)}_{\epsilon+\omega/2}-
\frac{\bar{\Gamma}\sqrt{x}}{2(1+x)}[(2\nu\Delta^{(-)}_{\epsilon+\omega}-\nonumber\\
&&2i\vec{n}\vec{\tau}(xf_R(\epsilon+\omega/2)+\frac{1}{2}\Delta^{(+)}_{\epsilon+\omega/2})]\nonumber
\end{eqnarray}
\begin{eqnarray}
\delta Q^{12} (\omega)&=&\frac{\bar{\Gamma}\sqrt{x}}{1+x}\{i\vec{n}\vec{\tau}[1+x-
(f_R(\epsilon+\omega/2)\times\nonumber\\
&& (f_L(\epsilon-\omega/2)+xf_R(\epsilon-\omega/2))+(\omega\rightarrow -\omega))]\nonumber\\
 && -\nu[f_R(\epsilon+\omega/2)
 f_L(\epsilon-\omega/2)-(\omega\rightarrow -\omega)]\}\nonumber
\end{eqnarray}
These formulas for  $\delta Q(\omega)$ can be applied for all $x$ if  modifications which come from  $ D(\epsilon,\epsilon')$ and $ E(\epsilon,\epsilon')$ vertices are included.  $D(\epsilon,\epsilon')$ introduces a factor $1+T_B(3-x)/2\sqrt{R_B}$ into the first term in expressions for
$\delta Q^{11}(\omega)$ and $\delta Q^{22}(\omega)$. There is also a contribution to $\delta Q^{12}(\omega)$:
$i T_B\bar{\Gamma}(f_L(\epsilon+\omega/2)f_L(\epsilon-\omega/2)-(L\rightarrow R))$. All these additions do not influence the resonance part of the tunneling. If we consider the limit of large $x$ the vertex $ E(\epsilon,\epsilon')$ is important. In this case we can directly obtain that to main order in $x$ it councils all terms in  vertex $ \delta Q (\omega)$ which are responsible for resonant spin orbit scattering.
With the help of these vertex functions we calculate all parts of the $S_{d2}$ noise power (see Eq.(\ref{STD})):
 \begin{eqnarray}
  S_0 (\omega)&=& \frac{e^2\bar{\Gamma}^2}{16}R_B\int\frac{d\epsilon}{2\pi}\Tr(\hat{q}^{RR}+\hat{q}^{AA})\Delta^{(-)}_{\epsilon-\omega/2}\Delta^{(-)}_{\epsilon+\omega/2}
 \nonumber
 \end{eqnarray}
 \begin{eqnarray}
 S^K_1(\omega) &=& -\frac{1}{4}e^2\bar{\Gamma}^2\sqrt{R_B}\int\frac{d\epsilon}{2\pi}
   \Tr[(\hat{q}^{RK}-\hat{q}^{KA})
   \vec{n}\vec{\tau}\Delta^{(-)}_{\epsilon-\omega/2}+\nonumber\\
  && (\omega\rightarrow-\omega)]\label{K1}
  \end{eqnarray}
  \begin{eqnarray}
   S_1^{AA+RR}(\omega) &=& \frac{e^2\bar{\Gamma}^2 \sqrt{R_B}}{4(1+x)}
 \int\frac{d\epsilon}{2\pi}\Tr \{-i\nu(\hat{q}^{RR}-\nonumber\\
 &&\hat{q}^{AA})
 \Delta^{(-)}_{\epsilon-\omega/2}\Delta^{(-)}_{\epsilon+\omega/2}+\nonumber\\
 &&\frac{1}{4}(\hat{q}^{RR}+\hat{q}^{AA})\vec{n}\vec{\tau})[\Delta^{(-)}_{\epsilon-\omega/2}
 \Delta^{(+)}_{\epsilon-\omega/2}\nonumber\\
&& +(\omega \rightarrow -\omega)] \}\label{RR1}\nonumber
\end{eqnarray}
\begin{eqnarray}
  S_1(\omega)&=&S^K_1(\omega)+S_1^{AA+RR}(\omega)\nonumber
 \end{eqnarray}
 here
  \[\hat{q}^{ab}=G^a(\epsilon-\omega/2)G^b(\epsilon+\omega/2)\]
 and $a,b$  label the retarded, advanced or Keldysh GFs: $a(b)=R,A,K$.
 In (\ref{K1}) to order $\sqrt{x}$ we can take $G^K(\epsilon)=iIm G^R(\epsilon)\Delta^{(+)}_{\epsilon}$.
The linear in $x$ singular contribution $S_2$ of  $S_{d2}$  is presented in the main text Eq.(\ref{S2}).

\end{document}